\pdfoutput=1

\documentclass[11pt,twoside,a4paper,cmspaper,final,collab]{cms-tdr}
\def\svnVersion{473825P}\def\cmsCernNoTag{CERN-EP-2018-019}\def\cmsCernDate{\today}\def\cmsMessage{Published in the European Physical Journal C as \href{http://dx.doi.org/10.1140/epjc/s10052-018-6143-z}{\doi{10.1140/epjc/s10052-018-6143-z.}}}

\begin{document}\cmsNoteHeader{B2G-16-028}

\hyphenation{had-ron-i-za-tion}
\hyphenation{cal-or-i-me-ter}
\hyphenation{de-vices}
\RCS$HeadURL: svn+ssh://svn.cern.ch/reps/tdr2/papers/B2G-16-028/trunk/B2G-16-028.tex $
\RCS$Id: B2G-16-028.tex 468669 2018-07-15 19:54:49Z alverson $

\newlength\cmsFigWidth
\ifthenelse{\boolean{cms@external}}{\setlength\cmsFigWidth{0.85\columnwidth}}{\setlength\cmsFigWidth{0.4\textwidth}}
\ifthenelse{\boolean{cms@external}}{\providecommand{\cmsLeft}{top\xspace}}{\providecommand{\cmsLeft}{left\xspace}}
\ifthenelse{\boolean{cms@external}}{\providecommand{\cmsRight}{bottom\xspace}}{\providecommand{\cmsRight}{right\xspace}}

\cmsNoteHeader{B2G-16-015}

\providecommand{\NA}{\ensuremath{\text{---}}}
\ifthenelse{\boolean{cms@external}}{\providecommand{\CL}{C.L.\xspace}}{\providecommand{\CL}{CL\xspace}}

\newcommand{\LQt}{\ensuremath{\mathrm{LQ}_3}\xspace}
\newcommand{\ALQt}{\ensuremath{\overline{\mathrm{LQ}}_3}\xspace}
\newcommand{\pttop}{\ensuremath{p_{\mathrm{T}}^{\cPqt}}\xspace}

\newcommand{\CRA}{\ensuremath{\mathrm{CR}_{\mathrm{A}}}\xspace}
\newcommand{\CRBo}{\ensuremath{\mathrm{CR}_{\mathrm{B1}}}\xspace}
\newcommand{\CRBt}{\ensuremath{\mathrm{CR}_{\mathrm{B2}}}\xspace}

\newcommand{\ttbarf}{\ensuremath{\ttbar_{\mathrm{f}}}\xspace}
\newcommand{\ttbarpf}{\ensuremath{\ttbar_{\mathrm{p+f}}}\xspace}
\newcommand{\ttbarp}{\ensuremath{\ttbar_{\mathrm{p}}}\xspace}

\newcommand{\ST}{\ensuremath{S_{\mathrm{T}}}\xspace}

\newcommand{\Wjets}{\ensuremath{\text{\PW+jets}}\xspace}
\newcommand{\Zjets}{\ensuremath{\text{\PZ+jets}}\xspace}

\title{Search for third-generation scalar leptoquarks decaying to a top quark and a $\tau$ lepton at $\sqrt{s}=13\TeV$}

\date{\today}

\abstract{
A search for pair production of heavy scalar leptoquarks (LQs), each decaying
into a top quark and a $\tau$ lepton, is presented. The search considers final states
with an electron or a muon, one or two $\tau$ leptons that decayed to hadrons, and additional jets.
The data were collected in 2016 in proton-proton collisions at
$\sqrt{s}=13\TeV$ with the CMS detector at the LHC,
and correspond to an integrated luminosity of 35.9\fbinv.
No evidence for pair production of LQs is found.
Assuming a branching fraction of unity for the decay $\mathrm{LQ} \to \cPqt \tau$,
upper limits on the production cross section are set as a function of LQ mass,
excluding masses below 900\GeV at 95\% confidence level.
These results provide the most stringent limits to date on the production of scalar
LQs that decay to a top quark and a $\tau$ lepton.
}

\hypersetup{%
pdfauthor={CMS Collaboration},%
pdftitle={Search for third-generation scalar leptoquarks decaying to a top quark and a tau lepton at sqrt(s) = 13 TeV},%
pdfsubject={CMS},%
pdfkeywords={CMS, physics, B2G}}

\maketitle

\section{Introduction}
\label{sec:introduction}
{\tolerance=1200
Leptoquarks (LQs) are hypothetical particles that carry non-zero baryon and lepton quantum numbers.
They are charged under all standard model (SM) gauge groups, and their
possible quantum numbers can be restricted by the assumption that
their interactions with SM fermions are renormalizable and gauge
invariant~\cite{Buchmuller1987442}. The spin of an LQ state is either 0
(scalar LQ) or 1 (vector LQ).
Leptoquarks appear in theories beyond the SM
such as grand unified theories~\cite{PatiSalam, GeorgiGlashow, Fritzsch:1974nn},
technicolor models~\cite{Technicolor, Lane:1991qh} and other
compositeness scenarios~\cite{LightLeptoquarks, Gripaios:2009dq}, and
R-parity-violating (RPV) supersymmetric models~\cite{Farrar:1978xj, Barbier:2004ez}.
\par}

Third-generation scalar LQs (\LQt{}s) have recently received considerable theoretical interest,
as their existence can explain the anomaly in the $\PaB \to \PD \tau \PAGn$ and
$\PaB \to \PDast \tau \PAGn$ decay rates reported by the
BaBar~\cite{Lees:2012xj, Lees:2013uzd}, Belle~\cite{Matyja:2007kt, Bozek:2010xy, Huschle:2015rga},
and LHCb~\cite{Aaij:2015yra} Collaborations. These decay rates deviate from the SM predictions by about
four standard deviations~\cite{Dumont:2016xpj}, and
studies of the flavor structure of LQ couplings reveal that large couplings
to third-generation quarks and leptons could explain this
anomaly~\cite{Tanaka:2012nw, Sakaki:2013bfa, Dorsner:2013tla, Gripaios:2014tna}.
Third-generation LQs can appear in models in which only third-generation
quarks and leptons are unified~\cite{Chakdar:2012vv, Chakdar:2013haa} and therefore their existence
is not constrained by proton decay experiments.
All models that predict LQs with masses at the TeV scale and sizable couplings to
top quarks and $\tau$ leptons can be probed by the CMS experiment at the CERN LHC.

\begin{figure}[h]
  \centering
  \includegraphics[width=0.22\textwidth]{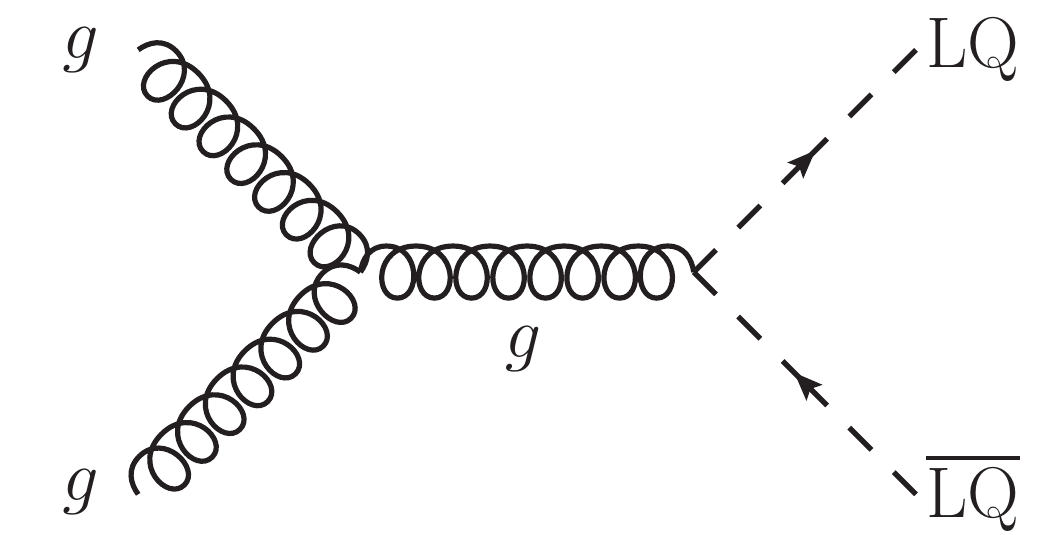} \qquad
  \includegraphics[width=0.22\textwidth]{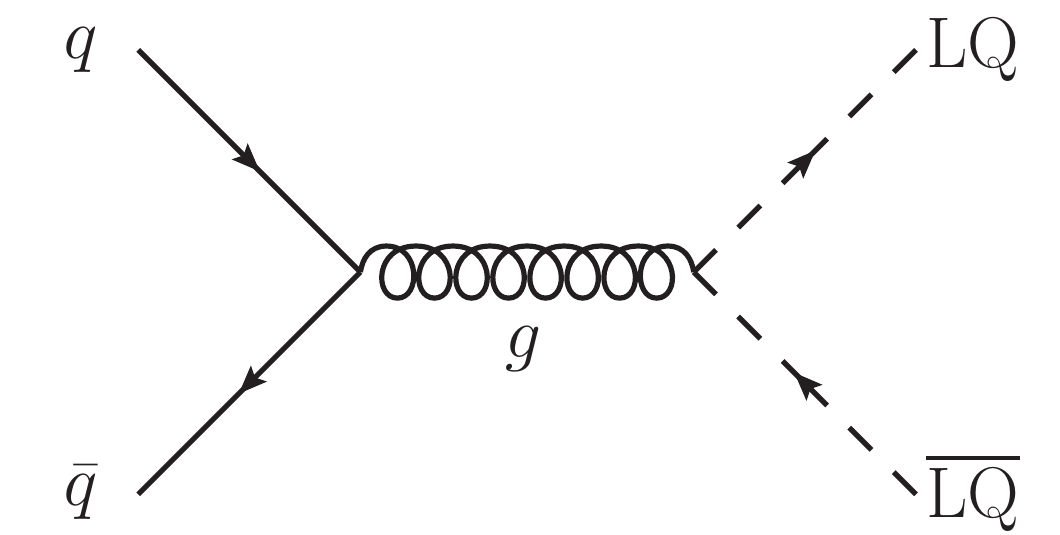} \\[0.2cm]
  \includegraphics[width=0.22\textwidth]{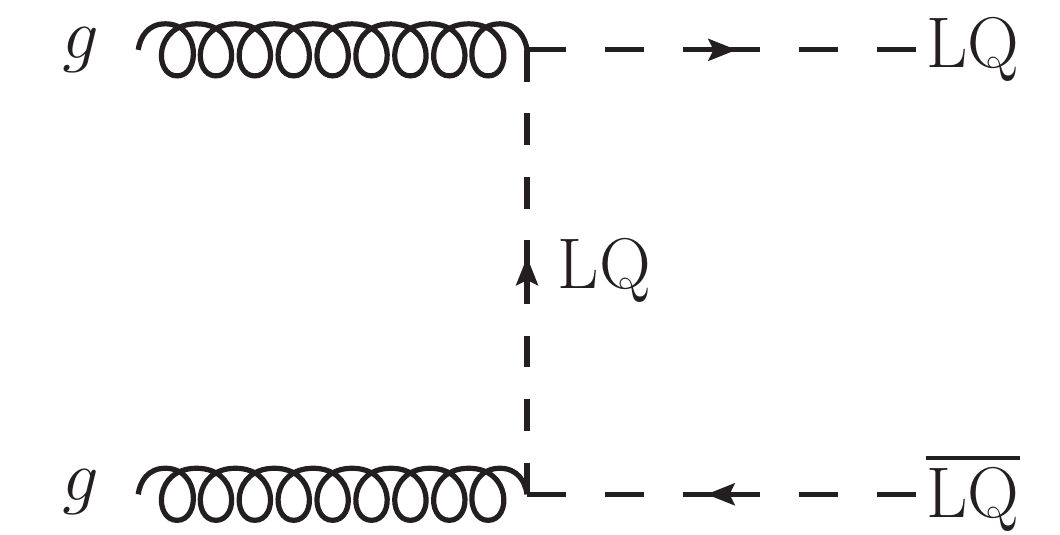} \qquad
  \includegraphics[width=0.22\textwidth]{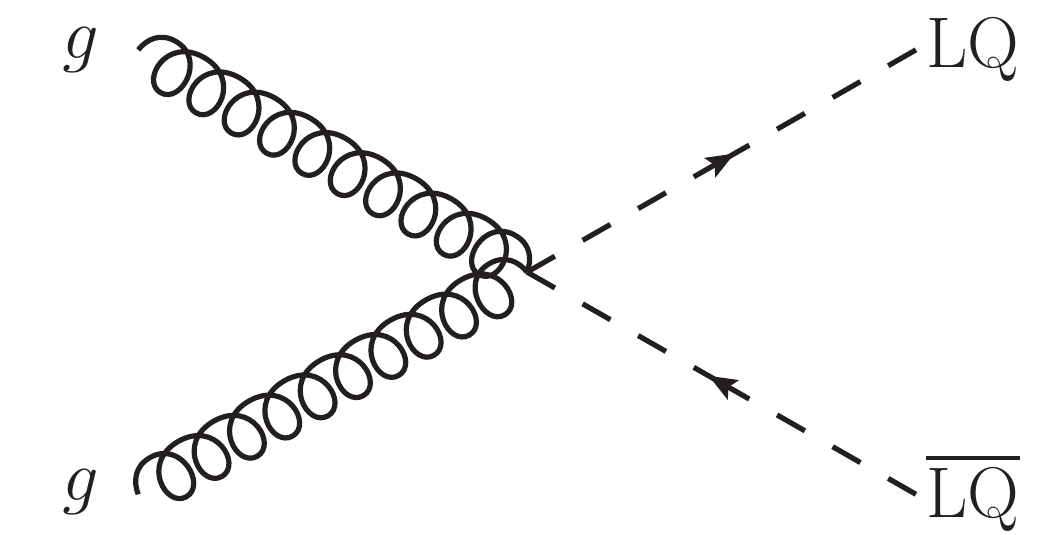} \\[0.2cm]
\caption{Dominant leading order Feynman diagrams for the production of leptoquark pairs
in proton-proton collisions.
\label{fig:LQpair_feynman}}
\end{figure}

In proton-proton ($\Pp\Pp$) collisions LQs are mainly pair produced through the quantum chromodynamic (QCD)
quark-antiquark annihilation and gluon-gluon fusion $s$- and $t$-channel sub\-pro\-ces\-ses as shown in
Fig.~\ref{fig:LQpair_feynman}.
There are also lepton-mediated $t$- and $u$-channel contributions that depend on the unknown
lepton-quark-LQ Yukawa coupling, but these contributions to \LQt production are negligible at the LHC
as they require third-generation quarks in the initial state.
Hence, the LQ pair-production cross section can be taken to depend only on the assumed values of the
LQ spin and mass, and on the center-of-mass energy.
The corresponding pair production cross sections have been calculated up to
next-to-leading order (NLO) in perturbative QCD~\cite{Kramer}.

This paper presents the first search for the production of an \LQt
decaying into a top quark and a $\tau$ lepton at $\sqrt{s} = 13\TeV$.
The search targets \LQt{}s with electric charges $-5/3\,e$ and $-1/3\,e$,
where $e$ is the proton charge, and with various possible weak isospin configurations,
depending on the model.
A previous search for this channel at $\sqrt{s} = 8\TeV$ by the
CMS Collaboration resulted in a lower mass limit of 685\GeV for an \LQt with branching
fraction $\mathcal{B}=1$ into a top quark and a $\tau$ lepton~\cite{Khachatryan:2015bsa}.
Other searches for an \LQt have targeted the decays $\LQt \to \cPqb \nu$ and
$\LQt \to \cPqb \tau$~\cite{Abazov:2007bsa, Aaltonen:2007rb, Aad:2013oea,
Chatrchyan:2012st, Khachatryan:2014ura, Khachatryan:2015wza, Aad:2015caa,
Aad:2015iea,Aaboud:2016nwl,Aaboud:2017wqg,
Khachatryan:2016jqo, Sirunyan:2017yrk,Sirunyan:2017kqq,Sirunyan:2017kiw}.
The results of the search presented here are also interpreted in the
context of RPV supersymmetric models,
where the supersymmetric partner of the bottom quark (bottom squark)
decays into a top quark and a $\tau$ lepton via the RPV coupling.

We consider events with at least one electron or muon and at least one $\tau$ lepton,
where the $\tau$ lepton undergoes a one- or three-prong hadronic decay,
$\tauh \to \text{hadron(s)}+\nu_\tau$.
In $\LQt\ALQt$ events, $\tau$ leptons arise directly from $\LQt$ decays, as well as from $\PW$
bosons in the top quark decay chain. Electrons and muons are produced in
leptonic decays of $\PW$ bosons or $\tau$ leptons.
Two search regions are used in this analysis: a di-$\tau$ region with the
signature $\ell\tauh\tauh$+jets and small background levels from SM processes, which provides
high sensitivity for \LQt masses below 500\GeV,
and a region with a single $\tau$ lepton in the final state, $\ell\tauh$+jets,
which has higher sensitivity for \LQt masses above 500\GeV because of a larger
signal efficiency.
Here, $\ell$ denotes either an electron or a muon.
The dominant backgrounds in this search come from $\ttbar$+jets
and $\Wjets$ production, with jets misidentified as hadronically decaying $\tau$ leptons.
These backgrounds are estimated through measurements in control regions and extrapolated to
the signal region.

In this paper, Section~\ref{sec:detector} describes the CMS detector, while Section~\ref{sec:simulation}
discusses the data samples and the properties of simulated events utilized in the analysis.
Section~\ref{sec:reconstruction} outlines the techniques used for event reconstruction and
Section~\ref{sec:selection} describes the selection criteria applied in each analysis channel.
The method used for the background estimation is reported in Section~\ref{sec:background}, and systematic
uncertainties are detailed in Section~\ref{sec:systematics}. Finally, Section~\ref{sec:results} contains
the results of the analysis,
and Section~\ref{sec:summary} summarizes this work.

\section{The CMS detector}
\label{sec:detector}

The central feature of the CMS apparatus~\cite{Chatrchyan:2008zzk}
is a superconducting solenoid of 6\unit{m} internal diameter,
providing a magnetic field of 3.8\unit{T}.
Within the solenoid volume are
a silicon pixel and strip tracker,
a lead tungstate crystal electromagnetic calorimeter (ECAL),
and a brass and scintillator hadron calorimeter (HCAL),
each composed of a barrel and two endcap sections.
Forward calorimeters extend the pseudorapidity ($\eta$) coverage
provided by the barrel and endcap detectors.
Electron momenta are estimated by combining the energy measurement
in the ECAL with the momentum measurement in the tracker.
Muons are measured in gas-ionization detectors
embedded in the steel flux-return yoke outside the solenoid.
A more detailed description of the CMS detector,
together with a definition of the coordinate system used and
the relevant kinematic variables, can be found in Ref.~\cite{Chatrchyan:2008zzk}.

Events of interest are selected using a two-tiered trigger system~\cite{Khachatryan:2016bia},
where the first level is composed of custom hardware processors and selects events at
a rate of around 100\unit{kHz} within a time interval of less than 4\mus.
The second level, known as the high-level trigger, uses a version of the full
event reconstruction software optimized for fast processing, and
reduces the event rate to around 1\unit{kHz} before data storage.

\section{Data sample and simulated events}
\label{sec:simulation}

The search for \LQt{}s presented here uses $\Pp\Pp$ collisions at $\sqrt{s}=13\TeV$ recorded with the CMS detector in 2016.
The data sample corresponds to an integrated luminosity of 35.9\fbinv~\cite{CMS-PAS-LUM-17-001}.

The leading order (LO) Monte Carlo (MC) program \PYTHIA~8.205~\cite{Sjostrand:2007gs}
is used to simulate the \LQt pair production signal process.
Both \LQt{}s are required to decay into a top quark and a $\tau$ lepton,
and polarization effects from the chiralities of the top quark and the $\tau$ lepton
have been neglected.
The signal samples are generated for \LQt masses ranging from 200 to 2000\GeV.

The principal background processes, top quark pair production (\ttbar) via the
strong interaction and electroweak single top quark production
in the $t$-channel and tW processes, are simulated with the NLO
generator \POWHEG (v1 is used for the single top \cPqt\PW\ processes and v2 for the single
top $t$-channel and \ttbar processes)~\cite{Nason:2004rx,Frixione:2007vw,
Alioli:2010xd,Frixione:2007nw,Alioli:2009je,Re:2010bp}.
The $s$-channel process of single top quark production is generated at NLO
using the program \MGvATNLO~(v2.2.2)~\cite{Alwall:2014hca}.
Other background processes involve \PW\ and \PZ\ boson production in association with jet radiation.
These processes are generated with \MGvATNLO~(v2.2.2), with
\PW\ boson production at NLO and \PZ\ boson production at LO level.
The matrix element generation of \PW\ and \PZ\ boson production is matched to the parton shower emissions with the Frederix and Frixione~\cite{Frederix:2012ps} and MLM~\cite{mlm} algorithms, respectively.
Background processes from QCD multijet production are simulated with \PYTHIA~8.205.
For all generated events, \PYTHIA~8.205 is used for the description of the parton shower and hadronization.
In the parton shower, the underlying event tune CUETP8M1~\cite{CMS-PAS-GEN-14-001,Skands:2014pea} has been applied for all samples except for
\ttbar and single top quark production in the $t$-channel, which use the underlying event tune CUETP8M2T4~\cite{CMS-PAS-GEN-14-001,Skands:2014pea}.
The event generation is performed using the NNPDF~3.0 parton distribution functions (PDFs)~\cite{Ball:2014uwa}, for all events.
The detector response is modeled with
the \GEANTfour~\cite{Agostinelli:2002hh} suite of programs.

\section{Event reconstruction}
\label{sec:reconstruction}

Event reconstruction is based on
the CMS particle-flow (PF) algorithm~\cite{Sirunyan:2017ulk},
which combines information from all subdetectors, including measurements
from the tracking system,
energy deposits in the ECAL and HCAL,
and tracks reconstructed in the muon detectors.
Based on this information,
all particles in the event are reconstructed as
electrons, muons, photons, charged hadrons, or neutral hadrons.

Interaction vertices are reconstructed using
a deterministic annealing filtering algorithm~\cite{Fruhwirth:99,Chatrchyan:2014fea}.
The reconstructed vertex with the largest value of summed physics-objects $\pt^2$ is taken
to be the primary $\Pp\Pp$ interaction vertex.
The physics objects are jets, clustered using the jet finding
algorithm~\cite{Cacciari:2008gp,Cacciari:2011ma}
with the tracks assigned to the vertex as inputs,
and the associated missing transverse momentum,
taken as the negative vector sum of the
\pt of those jets.
Charged particles associated with other interaction vertices
are removed from further consideration.

Muons are reconstructed using the information collected in the muon detectors
and the inner tracking detectors, and are
measured in the range $\abs{\eta}< 2.4$.
Tracks associated with muon candidates must be consistent with
muons originating from the primary vertex,
and are required to satisfy
a set of identification requirements.
Matching muon detector information to tracks measured in the silicon tracker results
in a \pt resolution for muons with
$20 <\pt < 100\GeV$ of 1.3--2.0\% in the barrel and
better than 6\% in the endcaps. The \pt resolution in the barrel
is better than 10\% for muons with \pt up to 1\TeV~\cite{Chatrchyan:2012xi}.

Electron candidates are reconstructed in the range $\abs{\eta}<2.5$
by combining tracking information with energy deposits in the ECAL.
Candidates are identified~\cite{electronreco} using information on the spatial
distribution of the shower,
the track quality and the spatial match between the track and electromagnetic cluster,
the fraction of total cluster energy in the HCAL, and
the level of activity in the surrounding tracker and calorimeter regions.
The transverse momentum \pt resolution for electrons with $\pt \approx 45\GeV$ from
$\Z \to \Pe \Pe$ decays ranges from 1.7\% for nonshowering electrons
in the barrel region to 4.5\% for electrons showering in the endcaps~\cite{electronreco}.

Jets are clustered using PF candidates as inputs to the anti-\kt algorithm~\cite{Cacciari:2008gp}
in the \FASTJET~3.0 software package~\cite{Cacciari:2011ma},
using a distance parameter of 0.4.
For all jets, corrections based on the jet area~\cite{Cacciari:2008gn}
are applied to the energy of the jets
to remove the energy contributions
from neutral hadrons from additional pp interactions in the same or adjacent bunch crossings (pileup collisions).
Subsequent corrections are used to account
for the nonlinear calorimetric response in both jet energy and mass,
as a function of $\eta$ and $\pt$~\cite{Chatrchyan:2011ds}.
The jet energy resolution amounts typically
to 15\% at 10\GeV, 8\% at 100\GeV, and 4\% at 1\TeV~\cite{Khachatryan:2016kdb}.
Corrections to the jet energy scale and the jet energy resolution
are propagated to the determination of the missing transverse momentum~\cite{Khachatryan:2016kdb}.
Jets associated with \cPqb{} quarks are identified
using the combined secondary vertex v2 algorithm~\cite{Chatrchyan:2012jua, Sirunyan:2017ezt}.
The working point used for jet \cPqb\ tagging in this analysis
has an efficiency of $\approx$65\% (in \ttbar simulated events)
and a mistag rate (the rate at which light-flavor jets are incorrectly tagged)
of approximately 1\%~\cite{Sirunyan:2017ezt}.

Hadronically decaying $\tau$ leptons are reconstructed with the hadron-plus-strips (HPS)
algorithm~\cite{Chatrchyan:2012zz} and are denoted by $\tauh$.
The HPS algorithm is based on PF jets and additionally includes photons originating from neutral pion decays.
Energy depositions in the ECAL are reconstructed in "strips" elongated in the direction of the azimuthal angle $\phi$, to take account of interactions in the material of the detector
and the axial magnetic field.
These deposits are associated with one or three charged tracks to
reconstruct various hadronic decay modes of $\tau$ leptons.
To suppress backgrounds from light-quark or gluon jets, a $\tauh$ candidate
is required to be isolated from other energy deposits in the event. The isolation criterion
is based on the scalar $\pt$ sum $I_\tau$ of charged and neutral PF candidates
within a cone of radius $\smash[b]{\sqrt{(\Delta \eta)^2 + (\Delta \phi)^2} = 0.5}$
around the $\tauh$ direction, excluding the $\tauh$
candidate. The isolation criterion is $I_\tau < 1.5\GeV$~\cite{CMS-PAS-TAU-16-002}.

The energies and resolutions as well as the selection efficiencies for all reconstructed
jets and leptons are studied in data and simulated events~\cite{Chatrchyan:2012xi, electronreco,
Khachatryan:2016kdb, Sirunyan:2017ezt, CMS-PAS-TAU-16-002}.
Based on these studies, the simulation is corrected to match the data.

\section{Event selection and categorization}
\label{sec:selection}

In the online trigger system, events with an isolated muon
(or electron) with $\pt>24\,(27)\GeV$ and $\abs{\eta}<2.4\,(2.1)$
are selected in the muon (electron) channel.
We select events offline containing
exactly one isolated muon (or electron) with $\pt>30\GeV$ and $\abs{\eta}<2.4\,(2.1)$.
For the electron channel, a veto is applied to events with a muon
to avoid overlap between the two channels.
At least one $\tauh$ lepton with $\pt>20\GeV$ and $\abs{\eta}<2.1$
and at least two jets with $\pt>50\GeV$ and $\abs{\eta} < 2.4$ are required.
Events are selected if a third jet with $\pt>30\GeV$ and $\abs{\eta} < 2.4$ is present,
and any additional jets are only considered if they have ${\pt>30\GeV}$.
The magnitude of the missing transverse momentum, \ptmiss, is required to be above 50\GeV.
Further, the events are divided into two categories corresponding to the number of
observed LQ candidates,
allowing the sensitivity to be enhanced over a broad range of LQ masses.
The event selection was chosen to maximize the expected significance of a possible LQ signal.
A summary of the selection criteria for both categories is given in Table~\ref{tab:Selection}
and described below.

\begin{table*}
\centering
\topcaption{Summary of selection criteria in event categories A ($\ell \tauh$ + jets) and B ($\ell \tauh \tauh$ + jets), where $\ell = \mu, \Pe$. In category A, the two subcategories, OS and SS, are defined by the charge of the $\ell\tau_h$ pair. The fit variable used in each category is also shown.
\label{tab:Selection}}
\renewcommand{\arraystretch}{1.2}
\begin{tabular}{l c c c}
 & \multicolumn{2}{c}{Category A} & Category B \\
 & OS $\ell \tauh$ + jets & SS $\ell \tauh$ + jets & OS $\ell\tau_h\tau_h$ + jets \\ \hline
Jet selection & $\geq$4 jets & $\geq$3 jets & $\geq$3 jets \\
\ptmiss selection & $\ptmiss>100\GeV$ & $\ptmiss>50\GeV$ & $\ptmiss>50\GeV$ \\
\tauh selection & \multicolumn{2}{c}{$\pt>100\GeV$} & $\pt^{\tau 1}>65\GeV$, $\pt^{\tau 2}>35\GeV$  \\
\cPqb{} tagging  & \multicolumn{2}{c}{$\geq$1 $\cPqb$ tag} & \NA \\
\ST selection & \multicolumn{2}{c}{\NA} & $\ST>350\GeV$ \\
Fit variable & \multicolumn{2}{c}{\pttop in two \ST bins}  & number of events \\
\end{tabular}
\end{table*}

\subsection{Category A: \texorpdfstring{$\ell \tauh$}{l tau} + jets}
{\tolerance=1200
In this category, exactly one $\tauh$ lepton is required
in addition to the presence of one electron or muon.
High \pt requirements are applied to
maximize the sensitivity at high LQ masses.
The leading jet is required to have ${\pt>150\GeV}$. In addition we define two subcategories
based on the electric charges
of the particles in the $\ell\tauh$ pair: opposite-sign (OS) and same-sign (SS).
Events passing the OS $\ell\tauh$ pair requirement must
contain at least four jets and have $\ptmiss>100\GeV$.
For both subcategories, we require that the leading tau
lepton has $\pt>100\GeV$ and that there is at least one $\cPqb$-tagged jet.
Finally the events are divided into two regions of \ST,
where \ST is the scalar $\pt$ sum of all selected jets, leptons, and $\ptmiss$.
In the low\,(high)-\ST search regions, events must satisfy $\ST<1200\,(\geq 1200)\GeV$.
This division adds sensitivity for \LQt masses of 600\GeV and higher.
\par}

The top quarks originating from the decay of a heavy \LQt are expected to be produced with
larger $\pt$ than the top quarks produced in background processes.
Therefore, the transverse momentum distribution of the top quark candidate decaying into hadronic jets (\pttop)
gives discrimination power between background and signal events, and
a measurement of the \pttop spectrum is performed in category~A.

A kinematic reconstruction of the top quark candidate is performed
by building top quark hypotheses using between one and five jets.
Because of the presence of multiple hypotheses in each event, we choose
the hypothesis in which the reconstructed top quark mass is closest to
the value of 172.5\GeV.

The statistical evaluation in this category is performed through a
template-based fit to the measured $\pttop$ distribution.

\subsection{Category B: \texorpdfstring{$\ell \tauh \tauh$}{l tau tau} + jets }
In this category events are required to have at least two \tauh leptons
and one electron or muon.
This requirement of two \tauh leptons removes a large fraction of the SM background processes.
The exception to this exclusion of SM backgrounds are diboson production events that contain one or more \tauh leptons,
but the cross sections for these processes are small.
The selection criteria in this category are adapted to provide good sensitivity for low LQ
masses.

Each event is required to contain an OS $\tauh\tauh$ pair.
If the event contains more than one $\tauh\tauh$ pair, the OS pair with the largest scalar \pt sum is selected.
Moreover, the leading and subleading $\tauh$ must satisfy $\pt>65\GeV$ and $35\GeV$,
respectively.

In this category a counting experiment is performed, as the number of expected background events is
too small for results to benefit from a shape-based analysis.

\section{Background estimation}
\label{sec:background}

The background in this analysis consists of samples of events that are selected
because of jets misidentified as $\tauh$ leptons and events with one electron or muon together with one or more \tauh leptons.

In the following, events from $\ttbar$ and \Wjets production that
contain at least one misidentified $\tauh$ lepton are obtained from control regions (CRs)
separately defined for the two search regions (SRs) A and B.
We consider the following contributions:
the $\ttbar$ background that consists of only misidentified $\tauh$ leptons
(or exactly one misidentified $\tauh$ lepton as in category A), denoted by $\ttbarf$,
the $\ttbar$ background that consists of (at least) one $\tauh$ lepton and (at least)
one misidentified $\tauh$ lepton (only used in category B), denoted by $\ttbarpf$, and
the $\ttbar$ background that consists of one $\tauh$ lepton,
denoted by $\ttbarp$.

An extrapolation method is used to derive the background due to misidentified $\tauh$ leptons.
The normalization, and in category A also the shape, of the $\ttbar$ background is
estimated using
\begin{equation}
N^{\ttbar \text{,\,data}}_{\text{SR}} = \left( N^{\text{data}}_{\text{CR}}-N^{\text{other,\,MC}}_{\text{CR}} \right) \, \frac{N^{\ttbar \text{,\,MC}}_{\text{SR}}}{N^{\ttbar \text{,\,MC}}_{\text{CR}}},
\label{eq:extrapolationfactor}
\end{equation}
where $N$ is the total number of events for the respective process in the signal region
or control region
and where ``other'' denotes all non-\ttbar background processes that are estimated from simulation.
The contribution to the background from events with $\tauh$ leptons only is estimated from simulated events.

\subsection{Backgrounds in category A}

In each subcategory of category A, the largest fraction of background events originates
from $\ttbar$ production.
The second largest source of background events arises from \Wjets production, while
minor contributions come from single top quark and \Zjets production.

The $\ttbarf$ background and the \Wjets background that contain a misidentified $\tauh$ lepton
are derived from a single control region (\CRA),
which is defined through the same selection requirements as for the SR,
but with an inverted isolation requirement for the $\tauh$ lepton.

\begin{figure}
  \centering
  \includegraphics[width=0.45\textwidth]{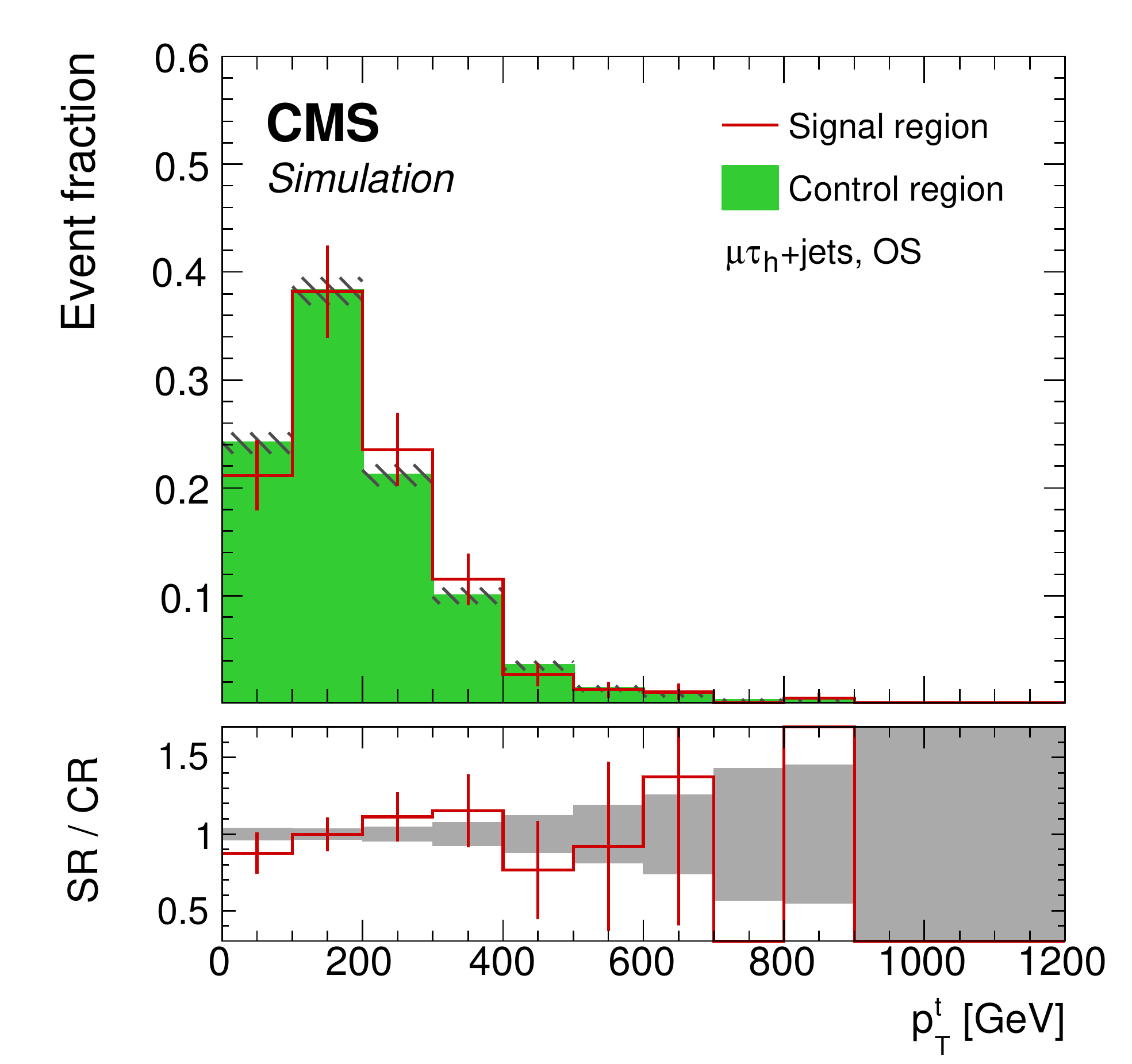} \qquad
  \includegraphics[width=0.45\textwidth]{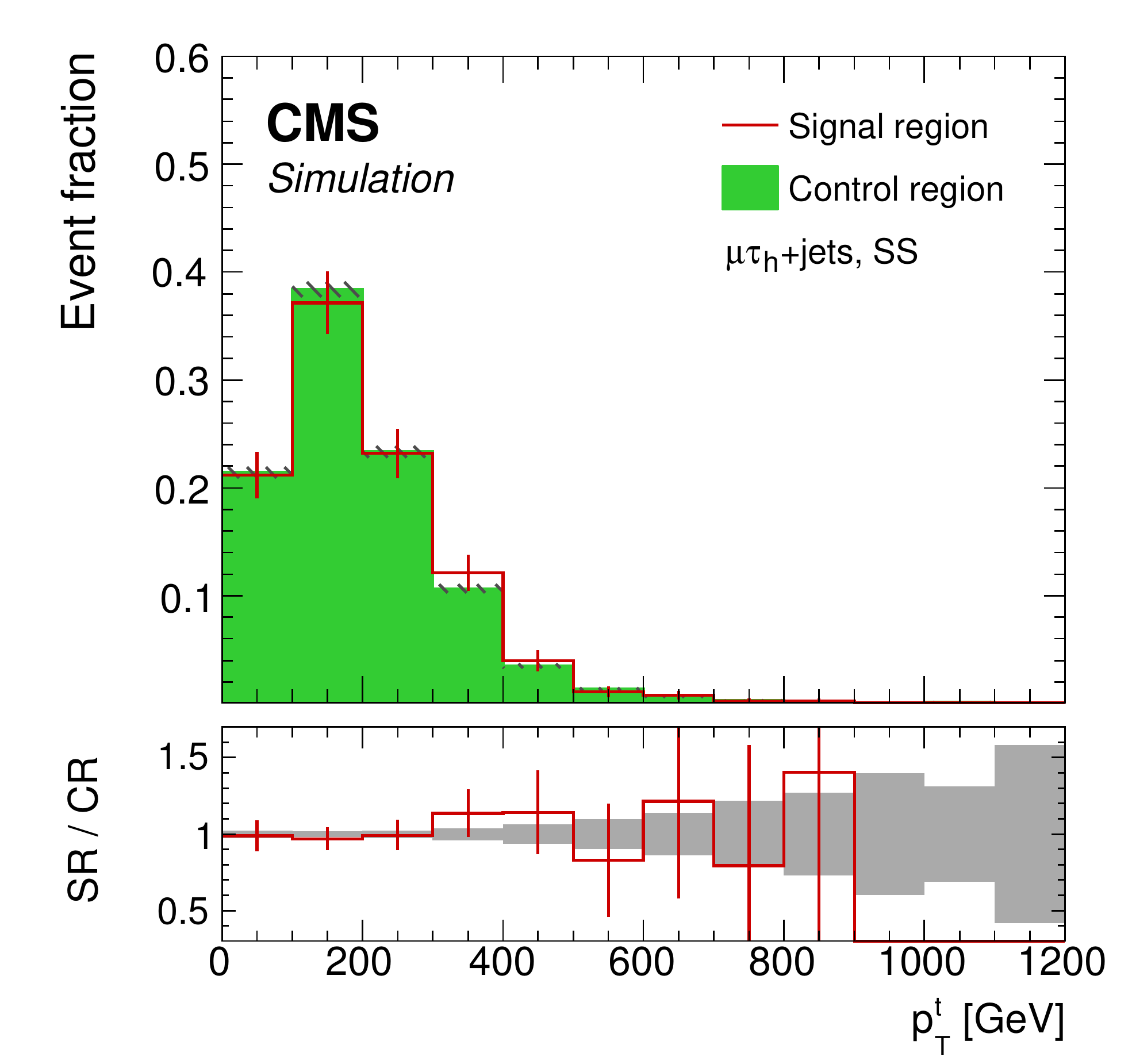} \qquad
  \caption{Shape comparison between the category A signal region
  and the corresponding control region, as a function of \pttop,
  for simulated \ttbar and $\Wjets$ events.
  Events with an opposite-sign $\mu \tauh$ pair are shown on the \cmsLeft, while those with a same-sign
  $\mu \tauh$ pair are shown on the \cmsRight. The full selection is applied and the $\ST$ categories are combined.
  All histograms are normalized to the total
  number of entries. Uncertainties of the signal region and control region are indicated by red error bars and gray hatched areas, respectively.
  The gray band in the ratio plot corresponds to the statistical uncertainty in the simulated samples.}
  \label{fig:shapecomparison}
\end{figure}

The shape of the $\pttop$ distribution is compared between the \CRA and
SR in simulated \ttbar and \Wjets events.
Since the inversion of the \tauh isolation criterion introduces kinematic differences
between the SRs and CRs, the
jet multiplicity and $\pttop$ are corrected in order to reproduce
the shape of the \ttbar and \Wjets backgrounds in the SRs~\cite{Stoever:2018},
as shown in Fig.~\ref{fig:shapecomparison}.

Once the kinematic distributions in the \CRA are corrected, we use Eq.~\eqref{eq:extrapolationfactor}
to extrapolate the \ttbar and \Wjets background yields to the SR.
In this equation, we replace $N^{\ttbar}$ with $N^{\ttbar,\,\Wjets}$ for category A.

\subsection{Backgrounds in category B}
\begin{figure}
  \centering
  \includegraphics[width=0.35\textwidth]{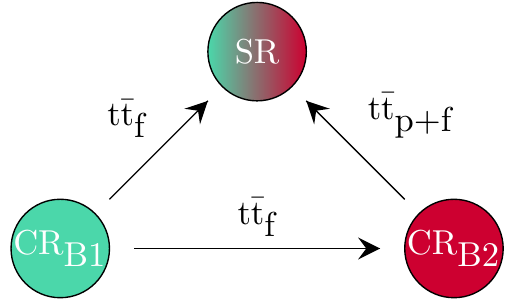}
  \caption{Strategy for the background estimation in category B. The $\ttbarf$ background in
  the signal region is derived from the control region \CRBo.
  The $\ttbarpf$ background in the signal region is derived from the control region
  \CRBt. To obtain an estimate of the $\ttbarf$ background in the control region
  \CRBt, the control region \CRBo is used.}
  \label{Fig:BGestimation_CatB}
\end{figure}
In category B, the dominant background also originates from $\ttbar$ production.
As the fraction of misidentified electrons and muons was found to be negligible
in this analysis, at least one of the two $\tauh$ leptons is mimicked by a jet.
Thus, background events from $\ttbar$ production consist either of only misidentified $\tauh$
leptons or one $\tauh$ lepton and one misidentified $\tauh$ lepton, plus an electron or a muon.
A separate CR is defined for each component.
The strategy for determining this background in category B
is shown in Fig.~\ref{Fig:BGestimation_CatB}.

The first control region (\CRBo) is defined by inverting the isolation criterion for all $\tauh$ leptons
with respect to the isolation criterion applied in the SR.
The region \CRBo is used to extrapolate the $\ttbarf$ background to the SR.
In contrast to the SR, the charge criterion on the $\tauh$
lepton is removed and the leading $\tauh$ lepton must have $\pt<100\GeV$ to avoid
overlap between the control region \CRBo and control region \CRA.
The $\ttbarf$ background normalization is then derived as in Eq.~\eqref{eq:extrapolationfactor}.

A second control region (\CRBt) to estimate the $\ttbarpf$ background is defined,
in which at least one isolated and at least one nonisolated $\tauh$ lepton are required.
In contrast to the SR, the charge criterion on the $\tauh$
lepton is removed and the leading $\tauh$ lepton must have $\pt<45\GeV$.
The event must have an opposite-sign $\ell\tauh$ pair.
For this requirement, the pair with the largest summed \pt is chosen.
In addition, the events must satisfy $M_{\mathrm{T}}(\ell,\ptmiss)>100\GeV$,
where $M_{\mathrm{T}}(\ell,\ptmiss)$ is the transverse mass of the
lepton-$\vec{p}_{\mathrm{T}}^{\,\text{miss}}$ system and defined as
\begin{equation*}
\smash[b]{M_{\mathrm{T}}(\ell,\ptmiss)=\sqrt{2\pt^\ell\ptmiss
\left(1-\cos[\Delta\phi(\vec{p}_{\mathrm{T}}^{\,\ell},\vec{p}_{\mathrm{T}}^{\,\text{miss}})]\right)
}}.
\end{equation*}
The largest non-$\ttbarpf$ fraction in control region \CRBt arises from the $\ttbarf$ events.
The estimate of this background is derived from the control region \CRBo and extrapolated
to the control region \CRBt by using the extrapolation method as in
Eq.~\eqref{eq:extrapolationfactor}.
Once the $\ttbarf$ background is estimated from \CRBo,
it is subtracted from \CRBt. The $\ttbarpf$ background is extrapolated
to the SR by using the extrapolation method as
in Eq.~\eqref{eq:extrapolationfactor}.

\section{Systematic uncertainties}
\label{sec:systematics}

Systematic uncertainties can affect both the overall normalization of background
components,
and the shapes of the $\pttop$ distributions for signal and background processes.
Uncertainties in the MC simulation are applied to all simulated events
used in the signal and in the various control regions.
For each systematic uncertainty, the background estimation procedure
described in Section \ref{sec:background} is repeated
to study the impact of the respective systematic variation on the final result of the analysis.
In the following, the systematic uncertainties applied to the analysis are summarized.

\begin{itemize}
	\item The uncertainty in the integrated luminosity measurement recorded with the CMS detector in the 2016 run at $\sqrt{s}=13\TeV$ is 2.5\%~\cite{CMS-PAS-LUM-17-001}.
	\item The following uncertainties in the normalization of the background processes are included:
		\begin{itemize}
			\item[--] 5.6\% in the $\ttbar$ production cross section~\cite{Khachatryan:2016kzg} for $\ttbar$ events that include $\tau$ leptons,
			\item[--] 10\% for single top quark~\cite{Kidonakis:2012rm,Chatrchyan:2014tua,Sirunyan:2016cdg}, W+jets, and Z+jets production~\cite{Chatrchyan:2014mua},
			\item[--] 20\% for diboson production~\cite{Campbell:2011bn,Gehrmann:2014fva,Khachatryan:2016tgp}.
		\end{itemize}
	\item The estimation of pileup effects is based on the total inelastic cross section. This cross section is determined to be 69.2\unit{mb}. The uncertainty is taken into account by varying the total inelastic cross section by 5\%~\cite{Chatrchyan:2012nj}.
	\item Simulated events are corrected for lepton identification, trigger, and isolation
	efficiencies. The corresponding scale factors are applied as functions of $\abs{\eta}$ and $\pt$.
	The systematic uncertainties due to these corrections are taken into account by varying each
	scale factor within its uncertainty.
	\item The scale factors for the jet energy scale and the jet energy resolution are determined as functions of $\abs{\eta}$ and $\pt$~\cite{Khachatryan:2016kdb}. The effect of the uncertainties in these scale factors are considered by varying the scale factors within their uncertainties. These variations are propagated to the measurement of the \ptmiss.
	\item Scale factors for the \cPqb\ tagging efficiencies are applied. These scale factors are measured as a function of the jet $\pt$~\cite{Sirunyan:2017ezt}. The corresponding uncertainty is taken into account by varying the scale factors within their uncertainties.
	\item Various uncertainties in the $\tau$ lepton reconstruction are considered.
	An uncertainty of 5\% in the $\tau$ lepton identification is applied, with an additional
	uncertainty of $0.2\,\pt/(1\TeV)$. An uncertainty of 3\% in the $\tau$ lepton energy
	scale is taken into account, and an uncertainty in the charge misidentification
	rate of 2\% is applied~\cite{CMS-PAS-TAU-16-002}.
	\item Parton distribution functions from the NNPDF 3.0 set are used to generate simulated events for both background and signal samples. The uncertainties in the PDFs are determined according to the procedure described in Ref.~\cite{Butterworth:2015oua}. The associated PDF uncertainties in the signal acceptance are estimated following the prescription for the
	LHC~\cite{Butterworth:2015oua}.
	\item We consider uncertainties in the renormalization ($\mu_{\mathrm{R}}$) and factorization ($\mu_{\mathrm{F}}$) scales by varying the respective scales, both simultaneously and independently, by  factors between 0.5 and 2.
	\item We apply an uncertainty in the background estimation method by varying the extrapolation factors
	for background processes without $\tau$ leptons within their uncertainties.
	An additional uncertainty due to the correction factors used to reweight events
	in control region \CRA is applied.
\end{itemize}

The systematic uncertainties with the largest effects on the most important background
processes and on the signal are summarized in Table~\ref{tab:systs}.
The most important background processes are the $\ttbarf$, $\ttbarf$ and $\Wjets$, and $\ttbarpf$ backgrounds derived from data, and the $\ttbarp$ background taken from simulation. Also shown is the systematic uncertainty associated with the signal produced by
an \LQt whose mass is 700\GeV.
The impact of the different sources of uncertainty varies for different processes.
The uncertainty due to the variation in the scales $\mu_{\mathrm{R}}$ and $\mu_{\mathrm{F}}$ has a large impact on the $\ttbarp$ background, and is derived from simulation.
The uncertainty in the $\tau$ lepton identification has the largest effect on the signal sample.
For the backgrounds derived from several CRs, the uncertainty in the extrapolation factor has the largest impact.

\begin{table*}
\centering
\topcaption{Summary of largest systematic uncertainties for the $\ttbarf$ (and \Wjets) and $\ttbarpf$ backgrounds derived from data, for the $\ttbarp$ background obtained from simulation and for a leptoquark signal with a mass of 700\GeV.
Shown are the ranges of uncertainties, which are dependent on the search regions and the lepton channel type.}
\label{tab:systs}
\begin{tabular}{l|ccc|ccc}
		& \multicolumn{3}{c|}{Category A} & \multicolumn{3}{c}{Category B}  \\
  Uncertainty & $\ttbarp$ & $\ttbarf$\,+\,\Wjets & \LQt & $\ttbarf$ & $\ttbarpf$ & \LQt \\ \hline
  Scales ($\mu_{\mathrm{F}}$, $\mu_{\mathrm{R}}$) & 26--42\%  & 1--7\%  & \NA & 5--7\% & 2--6\% & \NA \\
  $\tau$ ID & 8--9\% & 0--1\% & 9--11\% & 0\% & 5--6\% & 18--20\% \\
  Bkg. estimate &  \NA & 6--18\%  & \NA & 26--30\% & 30--38\% & \NA \\
\end{tabular}
\end{table*}

\section{Results}
\label{sec:results}

The results of all search categories in the electron and muon channels are combined in a
binned-likelihood fit.
A statistical template-based analysis, using the measured \pttop distributions in category A
and a counting experiment with the events measured in category B, is performed
by using the \textsc{Theta} software package~\cite{theta}.
Each systematic uncertainty discussed in Sec.~\ref{sec:systematics}
is accounted for by a nuisance parameter in the likelihood formation.

\begin{figure*}
  \centering
  \includegraphics[width=0.45\textwidth]{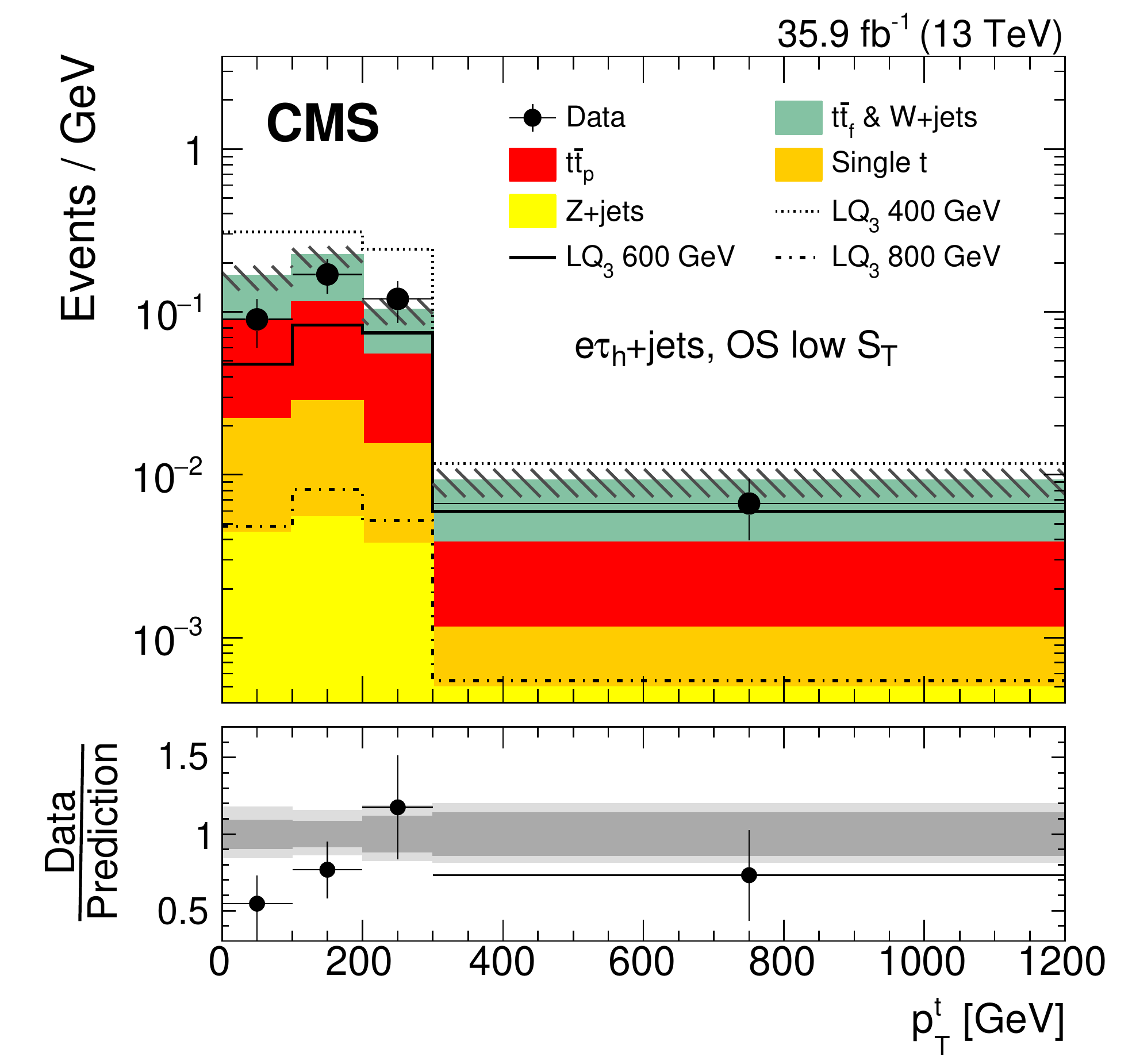}\qquad
  \includegraphics[width=0.45\textwidth]{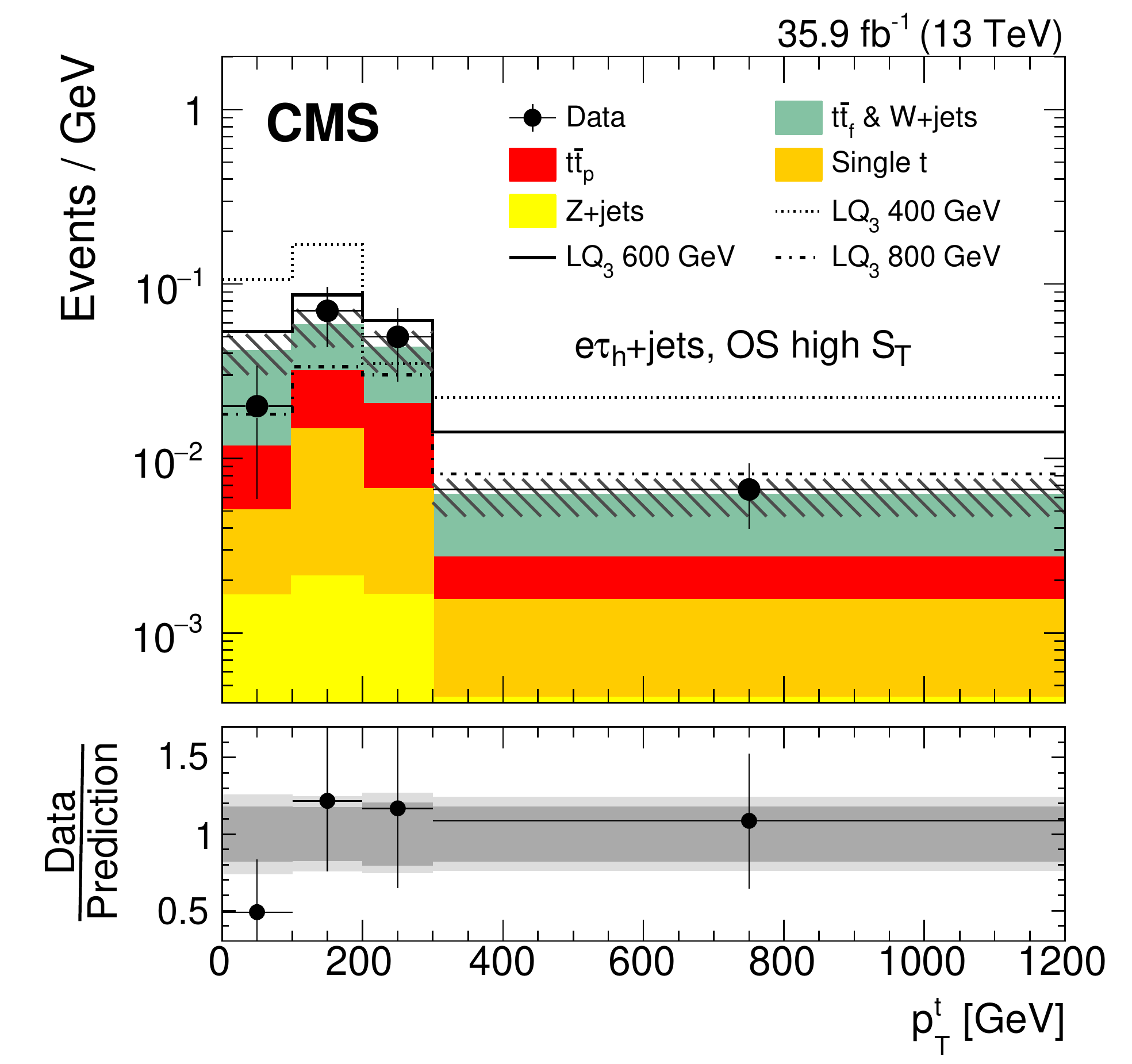} \\
  \includegraphics[width=0.45\textwidth]{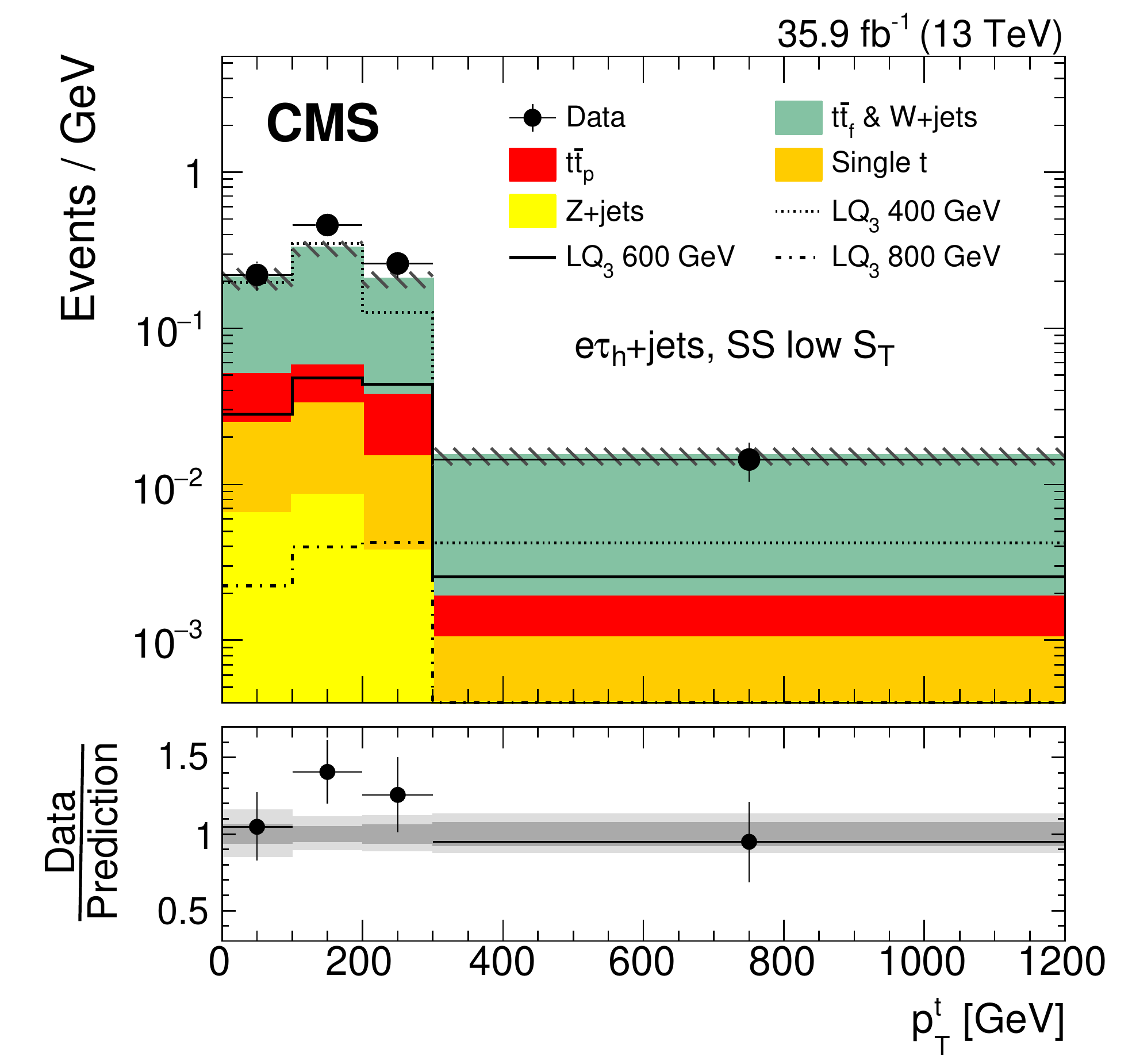}\qquad
  \includegraphics[width=0.45\textwidth]{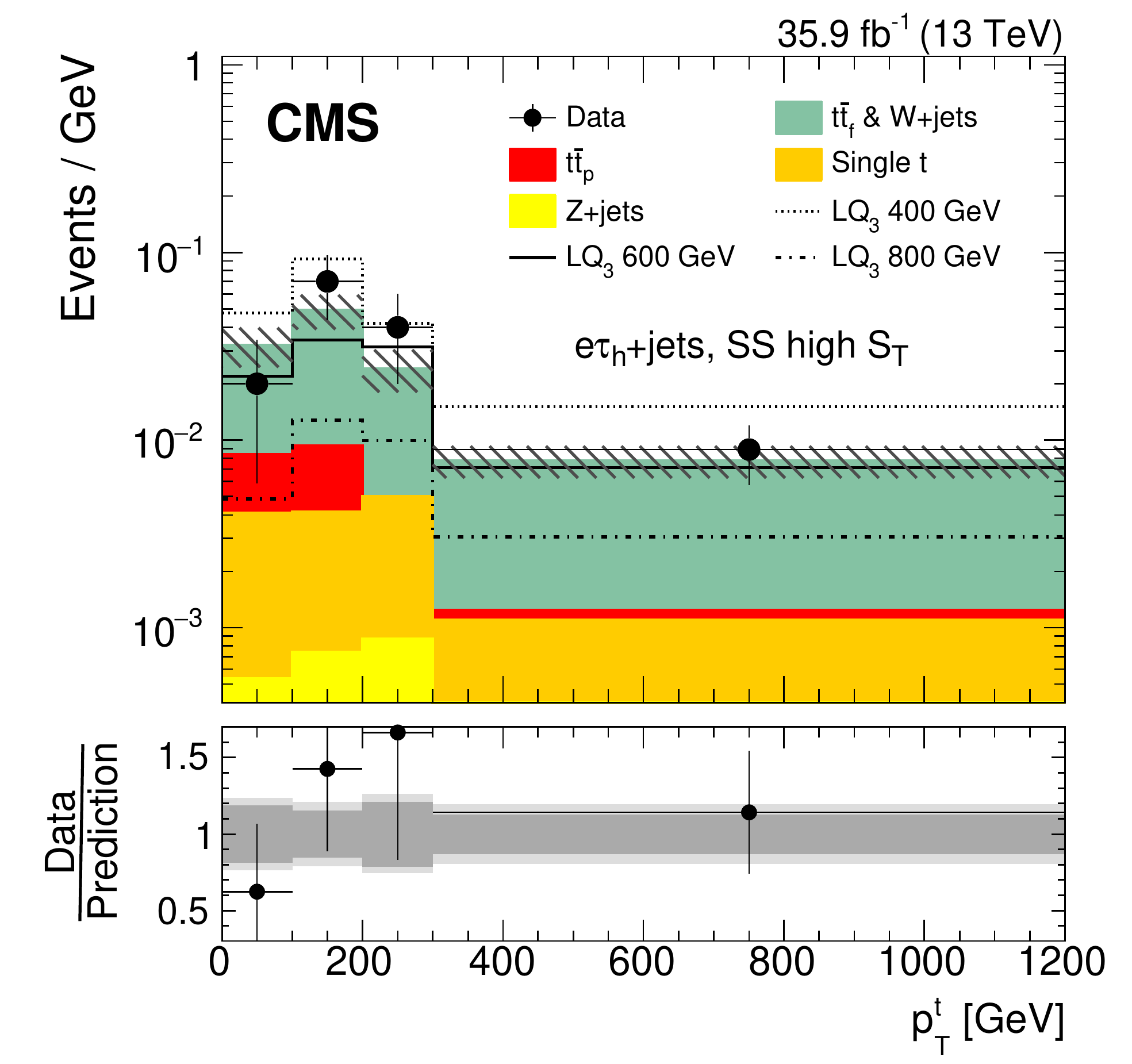}
  \caption{Distributions of \pttop for events in the electron channel passing the full selection in category A.
  The events are separated into OS (upper), SS (lower), low \ST (left) and
  high \ST (right) categories.
  The hatched areas represent the total uncertainties of the SM background.
  In the bottom panel, the ratio of data to SM background is shown together with
  statistical (dark gray) and total (light gray) uncertainties of the total SM background.}
  \label{fig:finalplots_electron}
\end{figure*}

\begin{figure*}
  \centering
  \includegraphics[width=0.45\textwidth]{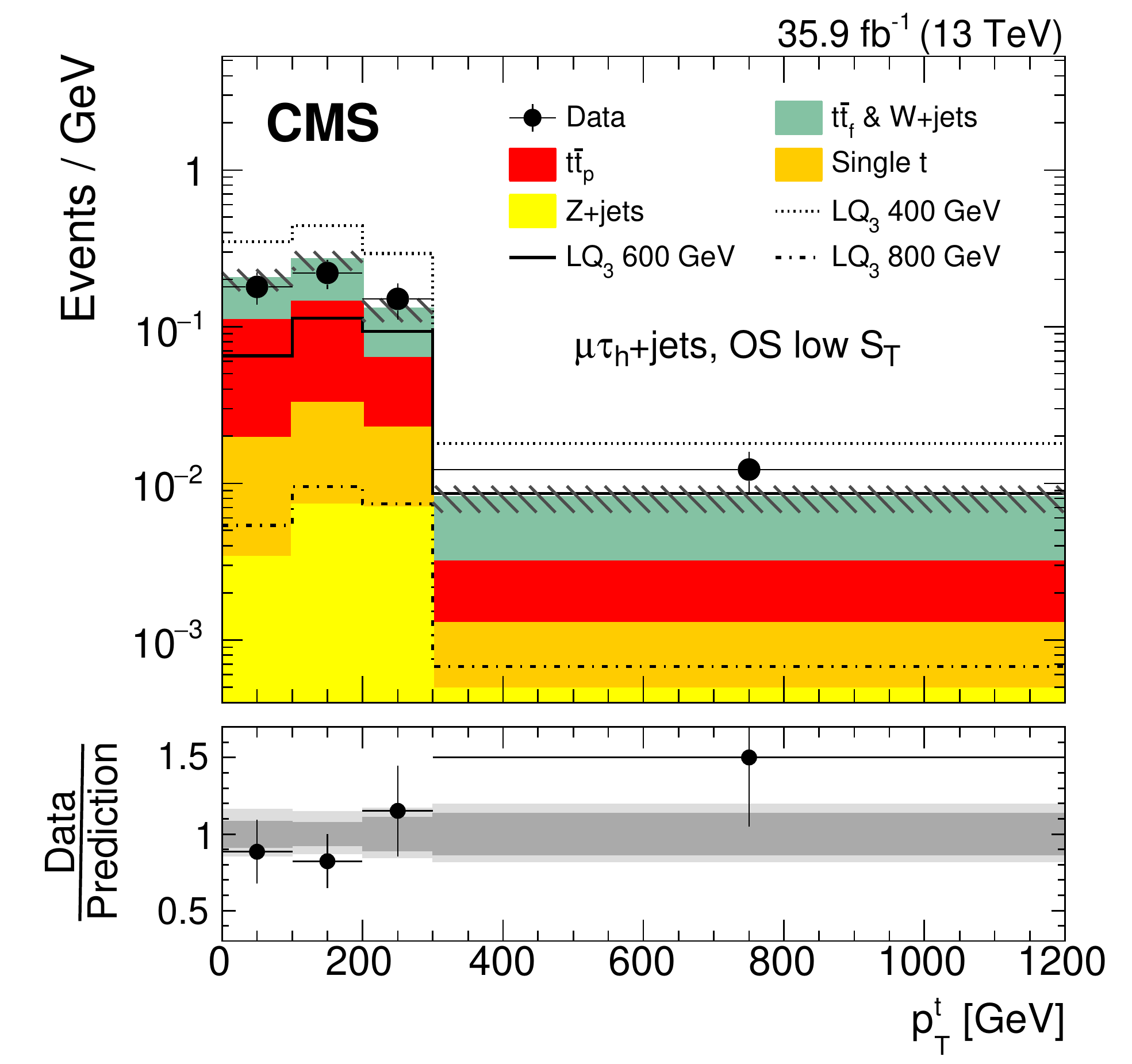} \qquad
  \includegraphics[width=0.45\textwidth]{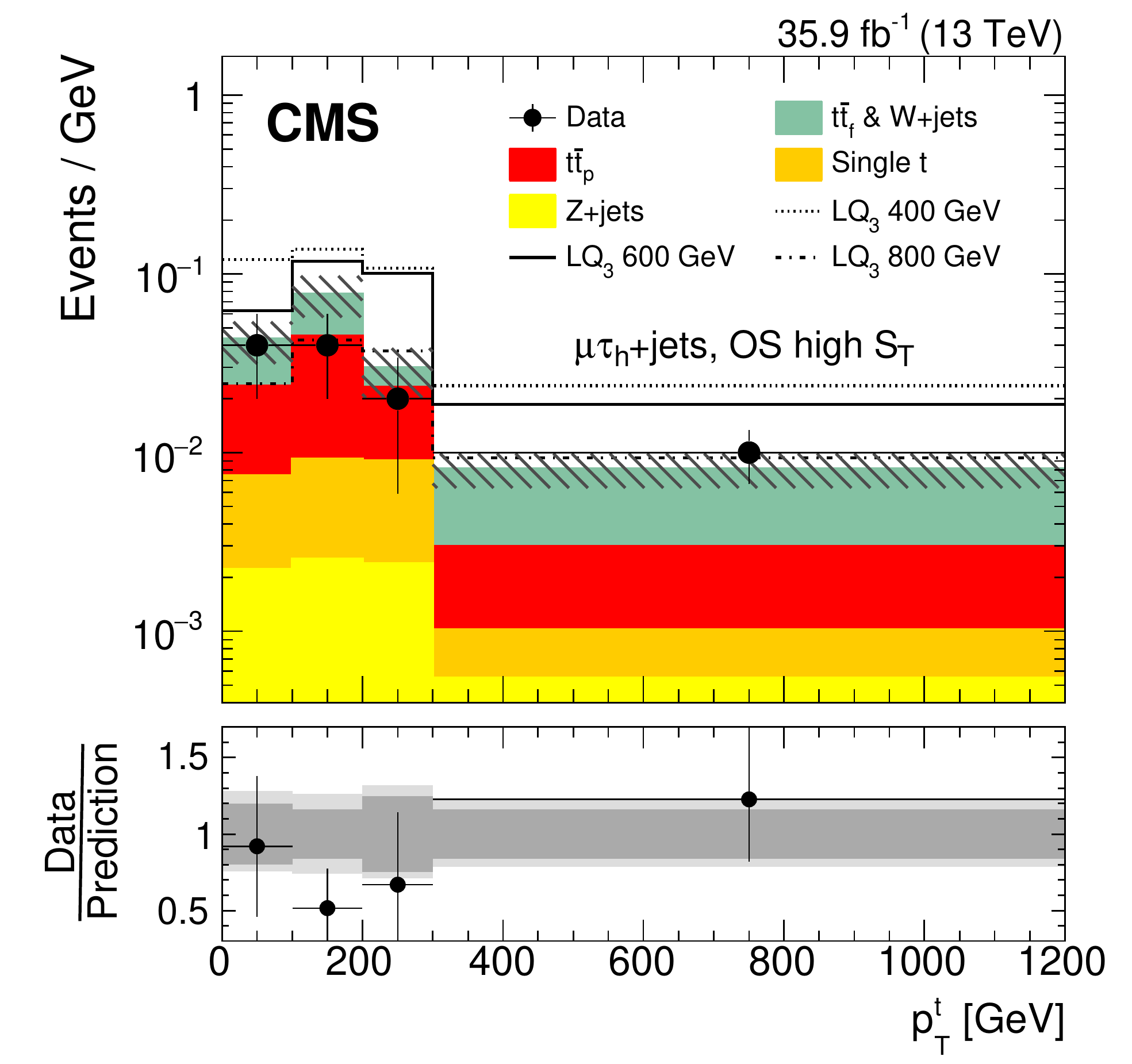} \\
  \includegraphics[width=0.45\textwidth]{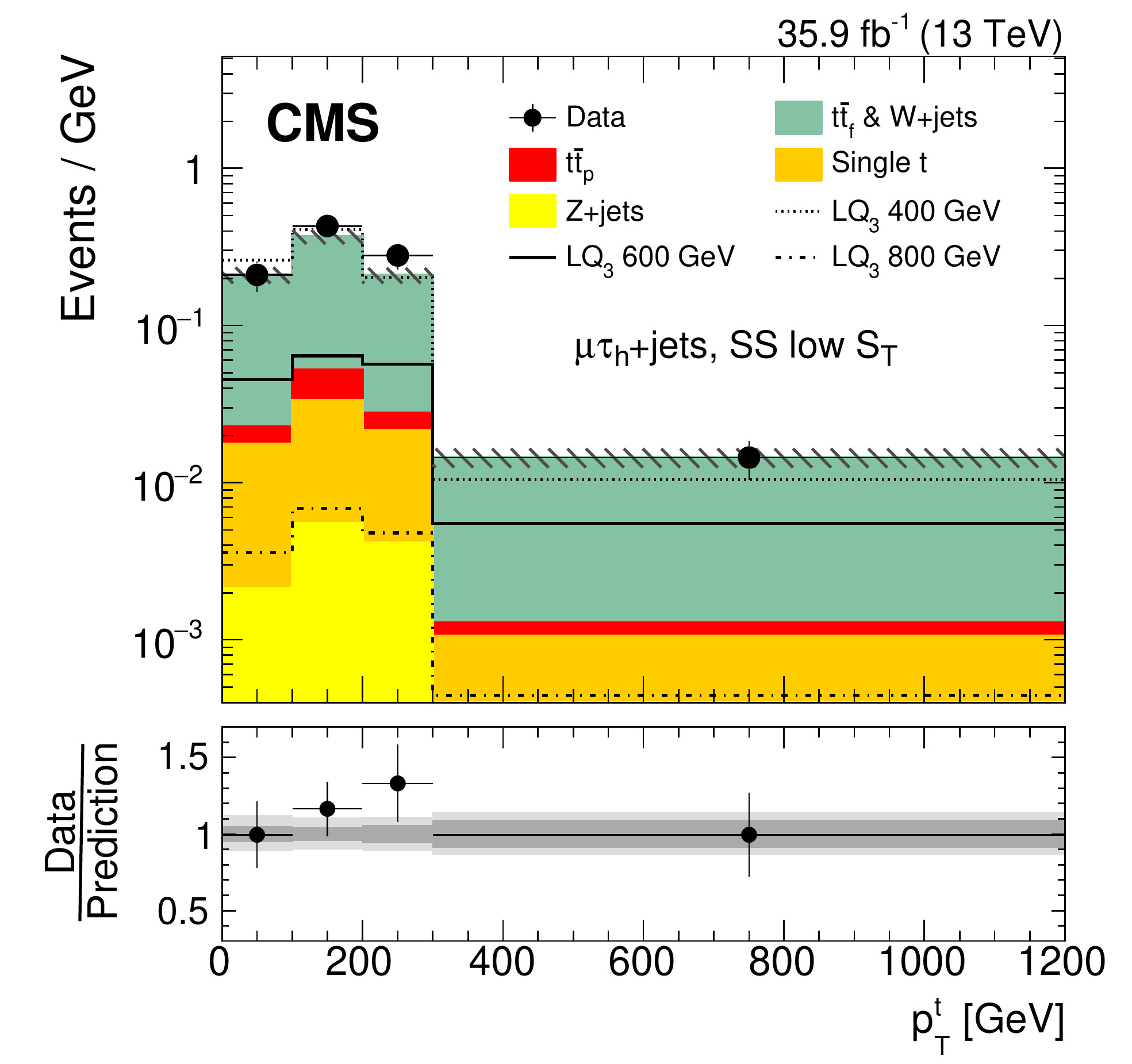} \qquad
  \includegraphics[width=0.45\textwidth]{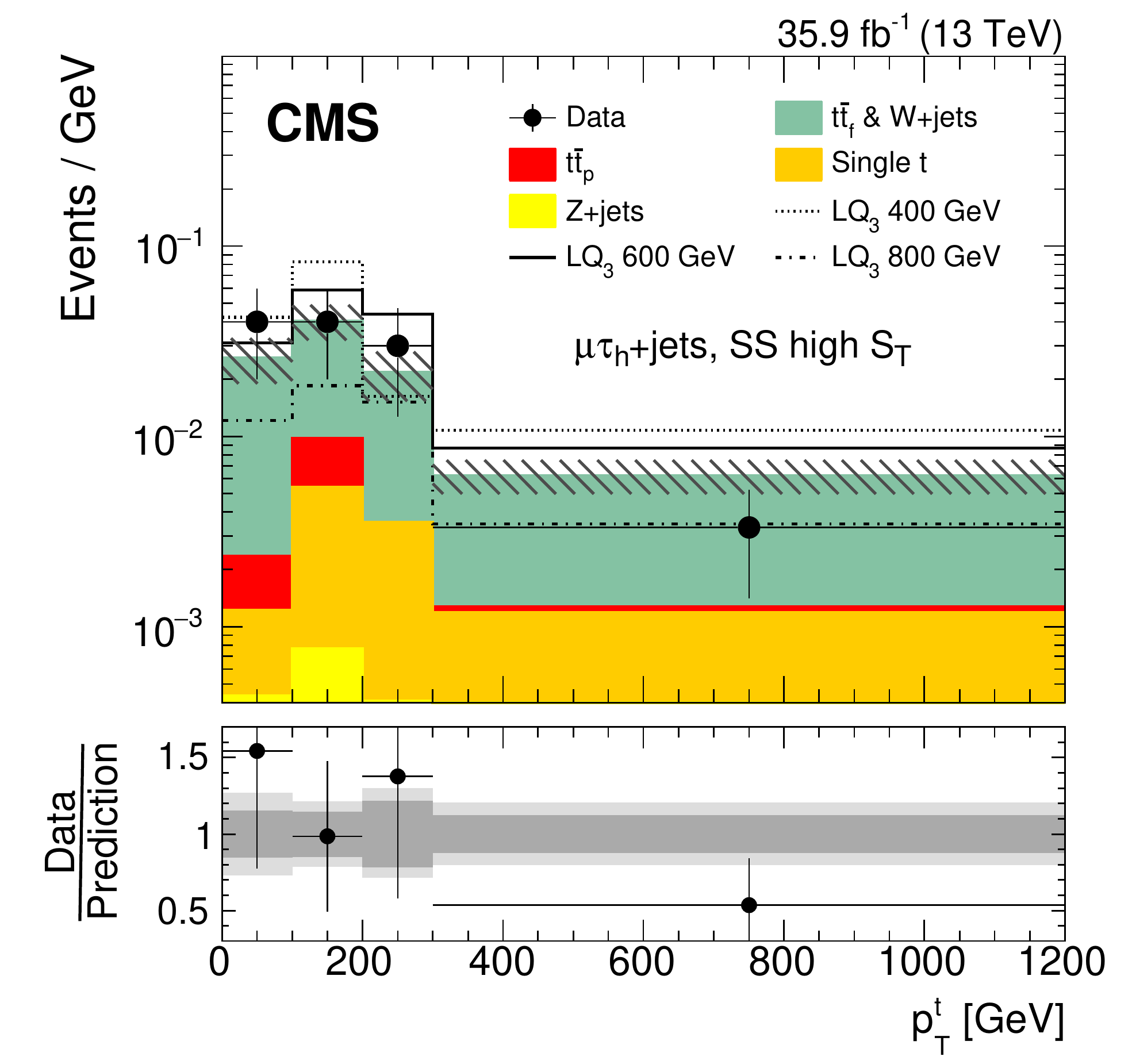}
  \caption{Distributions of \pttop for events in the muon channel passing the full selection in category A.
  The events are separated into OS (upper), SS (lower), low \ST (left) and
  high \ST (right) categories.
  The hatched areas represent the total uncertainties of the SM background.
  In the bottom panel, the ratio of data to SM background is shown together with
  statistical (dark gray) and total (light gray) uncertainties of the total SM background.
  \label{fig:finalplots_muon}}
\end{figure*}

The post-fit \pttop distributions in the electron and muon channels in category A
are shown in Figs.~\ref{fig:finalplots_electron} and \ref{fig:finalplots_muon},
respectively.
Contributions from \ttbar and \Wjets production with a misidentified \tauh
lepton are derived from control region \CRA, whereas SM backgrounds with a
\tauh lepton and other small backgrounds are taken from simulation.

\begin{table}
\topcaption{Final event yield in category B in the muon and electron channels for different
leptoquark mass hypotheses, the background processes, and data. The total uncertainties for the signal and the background processes are shown.
\label{tab:final_yields_categoryB}}
\centering
\renewcommand{\arraystretch}{1.2}
\begin{tabular}{@{\extracolsep{-0.3cm}} l  r l  r l }
Process          & \multicolumn{2}{c}{\quad$\Pe \tauh \tauh $ + jets\quad}      &
\multicolumn{2}{c}{\quad$\mu \tauh \tauh $ + jets\quad}     \\ \hline
\LQt (300\GeV)       & $ \qquad 97 $ & $^{+ 25 }_{- 24 }$   & $ \qquad 167  $ & $^{+ 36 }_{- 37 }$    \\
\LQt (400\GeV)       & $ 73 $ & $^{+ 14 }_{- 13 }$   & $ 98   $ & $^{+ 19 }_{- 17 }$    \\
\LQt (500\GeV)     & $ 34.1 $ & $^{+ 6.6 }_{- 6.2 }$ & $ 44.9 $ & $^{+ 8.5 }_{- 7.9 }$ \\
\LQt (600\GeV)     & $ 14.1 $ & $^{+ 2.8 }_{- 2.7 }$ & $ 21.1 $ & $^{+ 4.1 }_{- 3.8 }$ \\
\LQt (700\GeV)      & $ 7.3 $ & $^{+ 1.5 }_{- 1.4 }$ & $ 7.1  $ & $^{+ 1.5 }_{- 1.4 }$  \\
\LQt (800\GeV)      & $ 3.2 $ & $^{+ 0.7 }_{- 0.7 }$ & $ 4.4  $ & $^{+ 1.0 }_{- 0.9 }$  \\
\LQt (900\GeV)      & $ 1.5 $ & $^{+ 0.4 }_{- 0.3 }$ & $ 1.9  $ & $^{+ 0.4 }_{- 0.4 }$  \\
\LQt (1000\GeV)     & $ 0.8 $ & $^{+ 0.2 }_{- 0.2 }$ & $ 0.9  $ & $^{+ 0.2 }_{- 0.2 }$  \\
\hline
\ttbarf            & $ 2.5 $ & $ ^{+ 0.8 }_{- 1.2 }$ & $ 3.2 $ & $ ^{+ 1.5 }_{- 1.2 }$   \\
\ttbarpf           & $ 1.5 $ & $ ^{+ 0.8 }_{- 0.8 }$ & $ 2.0 $ & $ ^{+ 0.8 }_{- 0.9 }$   \\
Single t           & $ 0.3 $ & $ ^{+ 0.3 }_{- 0.3 }$ & $ 0.0 $ & $ ^{+ 0.2 }_{- 0.0 }$   \\
W+jets             & $ 0.5 $ & $ ^{+ 1.2 }_{- 0.5 }$ & $ 0.4 $ & $ ^{+ 0.7 }_{- 0.4 }$   \\
Z+jets             & $ 1.4 $ & $ ^{+ 0.5 }_{- 0.5 }$ & $ 1.0 $ & $ ^{+ 0.4 }_{- 0.4 }$   \\
Diboson            & $ 1.6 $ & $ ^{+ 1.7 }_{- 1.6 }$ & $ 1.7 $ & $ ^{+ 1.8 }_{- 1.7 }$   \\ \hline
Total background   & $ 7.9 $ & $ ^{+ 2.4 }_{- 2.5 }$ & $ 8.4 $ & $ ^{+ 2.6 }_{- 2.3 }$   \\ \hline
Data               & \multicolumn{2}{c}{9}       &    \multicolumn{2}{c}{11}           \\
\end{tabular}
\end{table}
In Table~\ref{tab:final_yields_categoryB}, the total number of events from background
processes and signal processes in category B is summarized.
No significant deviation from the SM prediction is observed in the data in either category A or category B.

A Bayesian statistical method~\cite{theta, bayesbook} is used to derive 95\% confidence level (\CL)
upper limits on the product of the cross section and the branching fraction squared for $\LQt$ pair production.
Pseudo-experiments are performed to extract expected limits under a background-only hypothesis.
For the signal cross section parameter, we use a uniform prior distribution.
For the nuisance parameters, log-normal prior distributions are used.
These are randomly varied within their ranges of validity to estimate the 68 and
95\% \CL expected limits.
Correlations between the systematic uncertainties across all channels are taken into account.
The statistical uncertainties of simulated samples are treated as an additional Poisson
nuisance parameter in each bin of the \pttop distribution.

\begin{figure}
\centering
\includegraphics[width=0.45\textwidth]{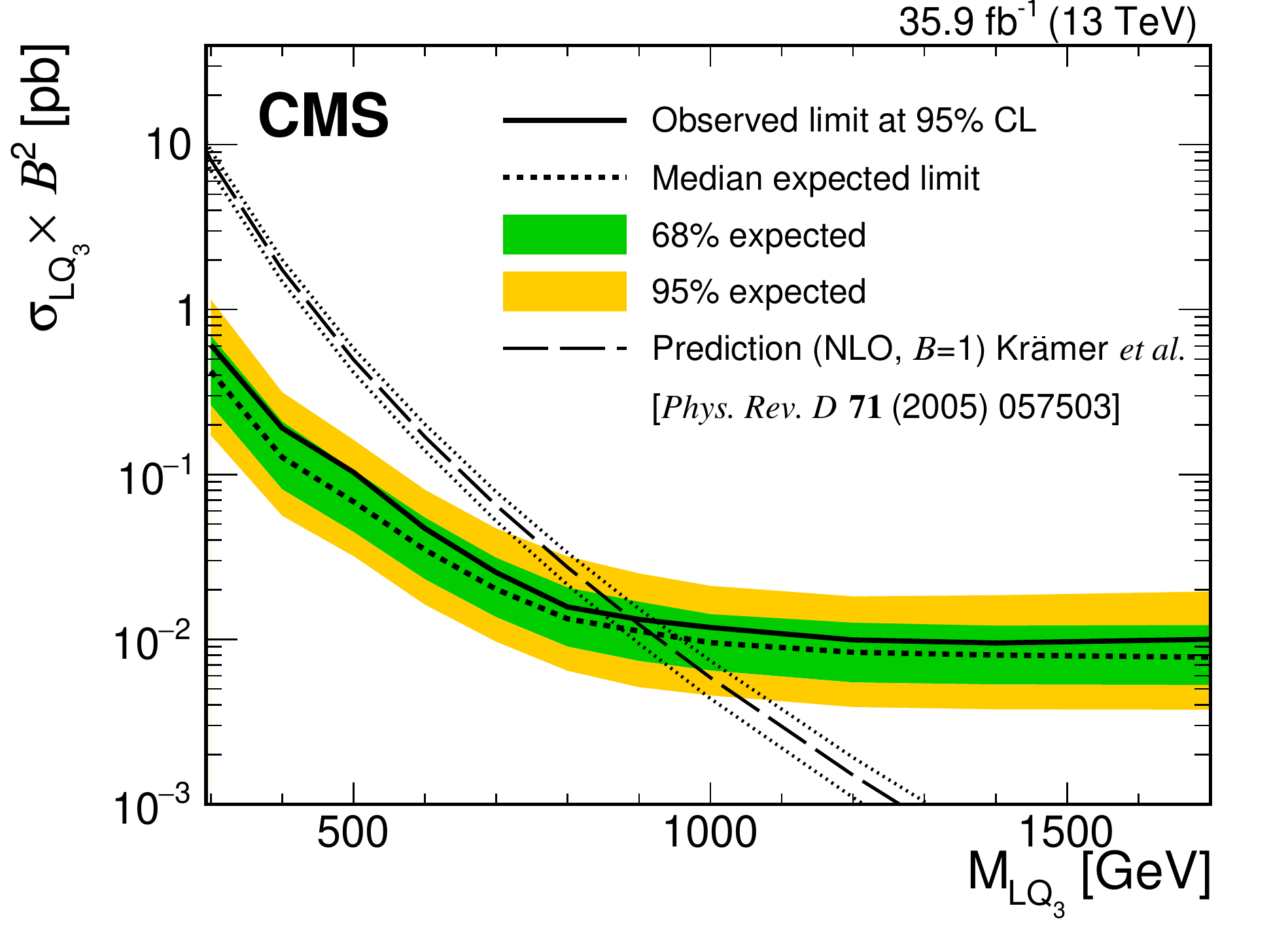}
\includegraphics[width=0.455\textwidth]{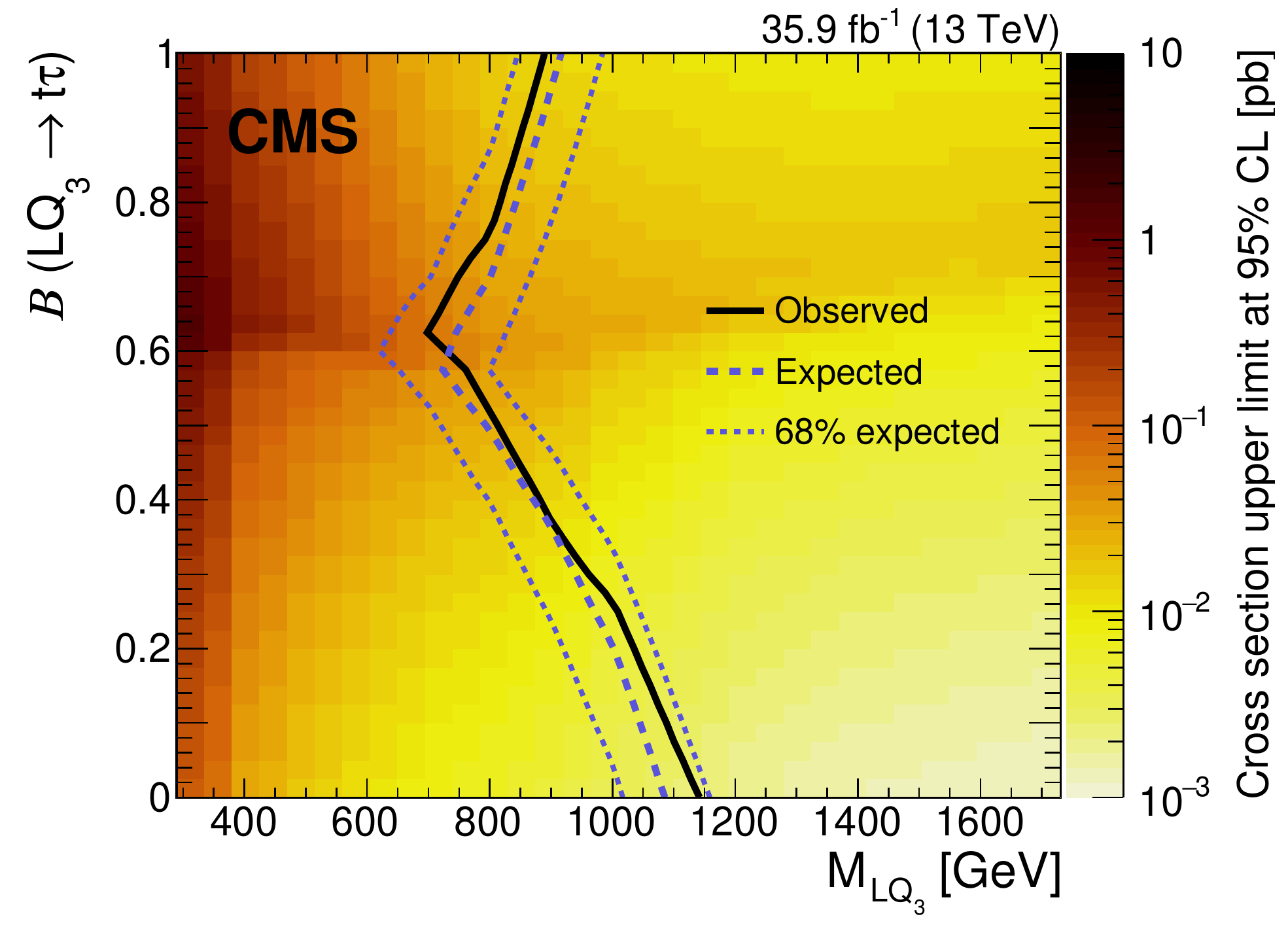}
\caption{Upper limits at 95\% confidence level on the product of the cross section and the branching fraction squared
(\cmsLeft), and on the leptoquark mass as a function of the branching fraction (\cmsRight),
for the pair production of scalar LQs decaying to a top quark and a $\tau$ lepton.
In the \cmsLeft plot, the theoretical curve corresponds to the NLO cross section with uncertainties from PDF and scale variations~\cite{Kramer}, shown by the dotted lines. The \cmsRight plot additionally includes results from a search for pair-produced bottom squarks~\cite{Sirunyan:2017kqq}.}
\label{fig:limitplot}
\end{figure}

The 95\% \CL upper limits on the product of the cross section and the branching
fraction squared $\mathcal{B}^2$
as a function of \LQt mass and the 95\% \CL upper limits on the \LQt mass as a function of $\mathcal{B}$
are shown in Fig.~\ref{fig:limitplot} (\cmsLeft).
The cross section for pair production of scalar LQs at NLO
accuracy~\cite{Kramer} is shown as the dashed line. The dotted
lines indicate the uncertainty due to the PDFs and to variations of the
renormalization and factorization scales by factors of 0.5 and 2.

Production cross sections of 0.6\unit{pb} for \LQt masses of 300\GeV and of about 0.01\unit{pb} for
masses up to 1.5\TeV are excluded at 95\% \CL under the assumption of
$\mathcal{B}=1$ for \LQt decays to a top quark and $\tau$ lepton.
Comparing these limits with the NLO cross sections,
\LQt masses up to 900\GeV (930\GeV expected) can be excluded.

Exclusion limits with varying branching fractions $\mathcal{B}$ are presented
in Fig.~\ref{fig:limitplot} (\cmsRight), where limits on the complementary
$\LQt\to \PQb\nu$ ($\mathcal{B}=0$) decay channel are also included.
The results for $\mathcal{B}=0$ are obtained from a search for pair-produced bottom
squarks~\cite{Sirunyan:2017kqq} with subsequent decays into $\PQb$ quark and neutralino pairs,
in the limit of vanishing neutralino masses.
Scalar \LQt{}s can be excluded for masses below 1150\GeV for $\mathcal{B}=0$ and for masses
below 700\GeV over the full $\mathcal{B}$ range.
For the assumptions of a LQ with symmetric couplings
under the SM gauge symmetry and with decays to only $\PQb\nu$ and $\cPqt\tau$,
$\mathcal{B}$ can only take values of 1 or 0.5. When these assumptions are lifted,
$\mathcal{B}$ can take all possible values between 0 and 1.
Note that if upper limits on $\mathcal{B}$ are to be used to constrain the
lepton-quark-\LQt Yukawa couplings,
$\lambda_{\PQb\nu}$ and $\lambda_{\PQt\tau}$, kinematic suppression factors that
favor $\PQb\nu$ decay over the $\PQt\tau$ decay have to be considered as
well~\cite{Abazov:2007bsa, Aaltonen:2007rb}.

The results presented here can be directly reinterpreted in the context of pair produced
down-type squarks decaying into top quark and $\tau$ lepton pairs.
Such squarks appear in RPV SUSY scenarios and correspond to LQs with
$\mathcal{B} = 0.5$.
These squarks are excluded up to a mass of 810\GeV, and the decay mode is dominated by the
RPV coupling $\lambda^{\prime}_{333}$~\cite{Dercks:2017lfq}.

\section{Summary}
\label{sec:summary}

A search has been conducted for pair production of third-generation scalar leptoquarks (\LQt{}s)
decaying into a top quark and a $\tau$ lepton. Proton-proton collision data
recorded in 2016 at a center-of-mass energy of 13\TeV,
corresponding to an integrated luminosity of 35.9\fbinv, has been analyzed.
The search has been carried out in the $\ell \tauh$+jets and $\ell \tauh \tauh$+jets channels,
where $\ell$ is either an electron or muon and $\tauh$ indicates a tau lepton decaying to hadrons.
Standard model backgrounds due to misidentified \tauh leptons are derived from control regions.
The measured transverse momentum distributions for the reconstructed top quark candidate
are analyzed in four search regions in the $\ell \tauh$+jets channel. The observed number of
events are found to be in agreement with the background predictions.

Upper limits on the production cross section of \LQt pairs are set between 0.6 and 0.01\unit{pb}
at 95\% confidence level for \LQt masses between 300 and 1700\GeV,
assuming a branching fraction of $\mathcal{B} = 1$.
The scalar \LQt{}s are excluded with masses below 900\GeV, for $\mathcal{B}=1$.
This result represents the most stringent limits to date on \LQt{}s
coupled to $\tau$ leptons and top quarks and constrains models
explaining flavor anomalies in the \cPqb\ quark sector through contributions from scalar LQs.

\begin{acknowledgments}
We congratulate our colleagues in the CERN accelerator departments for the excellent performance of the LHC and thank the technical and administrative staffs at CERN and at other CMS institutes for their contributions to the success of the CMS effort. In addition, we gratefully acknowledge the computing centers and personnel of the Worldwide LHC Computing Grid for delivering so effectively the computing infrastructure essential to our analyses. Finally, we acknowledge the enduring support for the construction and operation of the LHC and the CMS detector provided by the following funding agencies: BMWFW and FWF (Austria); FNRS and FWO (Belgium); CNPq, CAPES, FAPERJ, and FAPESP (Brazil); MES (Bulgaria); CERN; CAS, MoST, and NSFC (China); COLCIENCIAS (Colombia); MSES and CSF (Croatia); RPF (Cyprus); SENESCYT (Ecuador); MoER, ERC IUT, and ERDF (Estonia); Academy of Finland, MEC, and HIP (Finland); CEA and CNRS/IN2P3 (France); BMBF, DFG, and HGF (Germany); GSRT (Greece); NKFIA (Hungary); DAE and DST (India); IPM (Iran); SFI (Ireland); INFN (Italy); MSIP and NRF (Republic of Korea); LAS (Lithuania); MOE and UM (Malaysia); BUAP, CINVESTAV, CONACYT, LNS, SEP, and UASLP-FAI (Mexico); MBIE (New Zealand); PAEC (Pakistan); MSHE and NSC (Poland); FCT (Portugal); JINR (Dubna); MON, RosAtom, RAS, RFBR and RAEP (Russia); MESTD (Serbia); SEIDI, CPAN, PCTI and FEDER (Spain); Swiss Funding Agencies (Switzerland); MST (Taipei); ThEPCenter, IPST, STAR, and NSTDA (Thailand); TUBITAK and TAEK (Turkey); NASU and SFFR (Ukraine); STFC (United Kingdom); DOE and NSF (USA).

\hyphenation{Rachada-pisek} Individuals have received support from the Marie-Curie program and the European Research Council and Horizon 2020 Grant, contract No. 675440 (European Union); the Leventis Foundation; the A. P. Sloan Foundation; the Alexander von Humboldt Foundation; the Belgian Federal Science Policy Office; the Fonds pour la Formation \`a la Recherche dans l'Industrie et dans l'Agriculture (FRIA-Belgium); the Agentschap voor Innovatie door Wetenschap en Technologie (IWT-Belgium); the F.R.S.-FNRS and FWO (Belgium) under the ``Excellence of Science - EOS" - be.h project n. 30820817; the Ministry of Education, Youth and Sports (MEYS) of the Czech Republic; the Lend\"ulet ("Momentum") Program and the J\'anos Bolyai Research Scholarship of the Hungarian Academy of Sciences, the New National Excellence Program \'UNKP, the NKFIA research grants 123842, 123959, 124845, 124850 and 125105 (Hungary); the Council of Science and Industrial Research, India; the HOMING PLUS program of the Foundation for Polish Science, cofinanced from European Union, Regional Development Fund, the Mobility Plus program of the Ministry of Science and Higher Education, the National Science Center (Poland), contracts Harmonia 2014/14/M/ST2/00428, Opus 2014/13/B/ST2/02543, 2014/15/B/ST2/03998, and 2015/19/B/ST2/02861, Sonata-bis 2012/07/E/ST2/01406; the National Priorities Research Program by Qatar National Research Fund; the Programa Estatal de Fomento de la Investigaci{\'o}n Cient{\'i}fica y T{\'e}cnica de Excelencia Mar\'{\i}a de Maeztu, grant MDM-2015-0509 and the Programa Severo Ochoa del Principado de Asturias; the Thalis and Aristeia programs cofinanced by EU-ESF and the Greek NSRF; the Rachadapisek Sompot Fund for Postdoctoral Fellowship, Chulalongkorn University and the Chulalongkorn Academic into Its 2nd Century Project Advancement Project (Thailand); the Welch Foundation, contract C-1845; and the Weston Havens Foundation (USA).
\end{acknowledgments}

\bibliography{auto_generated}
\cleardoublepage \appendix\section{The CMS Collaboration \label{app:collab}}\begin{sloppypar}\hyphenpenalty=5000\widowpenalty=500\clubpenalty=5000\vskip\cmsinstskip
\textbf{Yerevan~Physics~Institute, Yerevan, Armenia}\\*[0pt]
A.M.~Sirunyan, A.~Tumasyan
\vskip\cmsinstskip
\textbf{Institut~f\"{u}r~Hochenergiephysik, Wien, Austria}\\*[0pt]
W.~Adam, F.~Ambrogi, E.~Asilar, T.~Bergauer, J.~Brandstetter, E.~Brondolin, M.~Dragicevic, J.~Er\"{o}, A.~Escalante~Del~Valle, M.~Flechl, M.~Friedl, R.~Fr\"{u}hwirth\cmsAuthorMark{1}, V.M.~Ghete, J.~Grossmann, J.~Hrubec, M.~Jeitler\cmsAuthorMark{1}, A.~K\"{o}nig, N.~Krammer, I.~Kr\"{a}tschmer, D.~Liko, T.~Madlener, I.~Mikulec, E.~Pree, N.~Rad, H.~Rohringer, J.~Schieck\cmsAuthorMark{1}, R.~Sch\"{o}fbeck, M.~Spanring, D.~Spitzbart, A.~Taurok, W.~Waltenberger, J.~Wittmann, C.-E.~Wulz\cmsAuthorMark{1}, M.~Zarucki
% ...
\vskip\cmsinstskip
% ...
1:~Also at~Vienna~University~of~Technology, Vienna, Austria\\
% ...
\end{sloppypar}
\end{document}